\documentclass[aps,singlecolumn,nofootinbib]{revtex4}
\usepackage{graphicx}
\usepackage{amsmath}
\usepackage{amssymb}
\usepackage{mathtools}
\usepackage{bm}

\DeclareMathOperator{\Ima}{im}

\newcommand{\rn}{{\rm n}}

\begin{document} 

\title{\bf Solving Moment Hierarchies for Chemical Reaction Networks} 

\author{Supriya Krishnamurthy}

\affiliation{Department of Physics, Stockholm University, SE- 106 91,
Stockholm, Sweden} 

\author{Eric Smith}

\affiliation{Earth-Life Science Institute, Tokyo Institute of
Technology, 2-12-1-IE-1 Ookayama, Meguro-ku, Tokyo 152-8550, Japan}

\affiliation{Department of Biology, Georgia Institute of
Technology, 310 Ferst Drive NW, Atlanta, GA 30332, USA}

\affiliation{Santa Fe Institute, 1399 Hyde Park Road, Santa Fe, NM
87501, USA}

\affiliation{Ronin Institute, 127 Haddon Place, Montclair, NJ 07043,
USA}

\date{\today}
\begin{abstract}
The study of Chemical Reaction Networks (CRN's) is a very active field.
Earlier well-known results \cite{Feinberg:def_01:87, Anderson:product_dist:10}
identify a topological
quantity called deficiency, for any CRN, which, when exactly equal to zero,
leads to a unique factorized steady-state for these networks.
No results exist however for the
steady states of non-zero-deficiency networks. In this paper, we show
how to write the full moment-hierarchy for any
non-zero-deficiency CRN obeying mass-action kinetics, in terms
of equations for the factorial moments.
Using these, we can
recursively predict values for lower moments from
higher moments, reversing the procedure usually
used to solve moment hierarchies. 
We show, for non-trivial examples, that in this manner
we can predict any moment of interest, for CRN's with non-zero deficiency
and non-factorizable steady states.
\end{abstract}

\maketitle

\section {Introduction}

Models of Chemical Reaction Networks (CRN's) are  ubiquitious
in the study of biochemistry, systems biology, ecology and epidemiology.
They provide a framework within which even several models
studied in physics, may be cast.
CRN's are defined
by a set of {\em species}, {\em complexes} and {\em reactions}
which, when taken together, specify the system of interest (see Figs.
\ref{fig:g_123_conn} and \ref{fig:two_species_bistable}).
The mathematical modeling of CRN's
is usually carried out either through the study of deterministic ODE's
(or {\it rate equations}) which specify the mean behaviour
of the concentrations of the different species, or by modeling
the stochastic variability of species counts
as a continuous-time Markov chain, where a transition occurs
every time a reaction takes place. One of the major results
pertaining to deterministic models of CRN's is the deficiency zero
theorem \cite{Feinberg:def_01:87, Horn:mass_action_72, Feinberg:notes:79}.
This relates a topological quantity
called the deficiency (a non-negative integer
index, denoted by $\delta$) to
the existence, uniqueness, and stability of positive fixed points
of the rate equations. In particular,
when $\delta= 0$ for CRN's which are weakly reversible \footnote {A
 weakly reversible CRN is one in which any complex can be transformed to any other, within one connected component of the network, via a directed path of reactions. A reversible network is one in which each reaction is accompanied by its reverse. Neither weakly reversible nor reversible networks  need to be {\em time-reversible}. So detailed balance does {\em not} generically hold.},
there is a unique, asymptotically stable equilibrium,
for any choice of (positive) rate constants.
A few theorems exist for deterministically-modeled
CRN's with $\delta > 0$ as well \cite{Feinberg:def_01:87,Craciun:multistability:05,Ji:PhD:11},
which either affirm that a given network is capable of multistationarity 
or can rule out this possibility (See \cite{Joshi:Survey:15}
for a recent survey).

Modeling CRN's by ODE's is however expected to be accurate
only when species concentrations are high. When this is not the case,
such as in, for example, gene expression \cite{Elowitz:stochastic:02, Ozbudak:single:02},
cell signaling \cite{Lestas:limits:10}
or  enzymatic processes \cite{Xie:enzyme:01},
then a stochastic modeling of CRN's is more appropriate.
A major result for this class of models is the theorem 
by Anderson, Craciun and Kurtz (ACK) \cite{Anderson:product_dist:10} (motivated by earlier work on queueing theory by Kelly \cite{Kelly:79}), who show
that if the conditions of the deficiency zero theorem hold
for a deterministically modeled CRN, then the corresponding stochastic system
has a product-form steady state.
There has also been work done \cite{Anderson:ACR:14} on the extinction time for certain reactions in stochastic models of CRN's of deficiency one.
However, no general results exist for obtaining the
steady-state behaviour of  CRN's with $\delta >0$.

This absence of general results reflects a deeper and more fundamental
feature of CRN's: unlike
simple random walks on ordinary graphs for which abundant results exist,
the elementary reaction events in CRN's involve \emph{concurrency}
in the conversion of inputs to outputs ~\cite{Danos:rule_based_modeling:08,Harmer:info_carriers:10}.  The underlying topology of a CRN is a \textit{multihypergraph} rather than an ordinary graph~\cite{Andersen:comp_rules:13,Andersen:generic_strat:14}; fewer results exist for hypergraphs because generic problems of search or optimization are computationally hard~\cite{Berge:hypergraphs:73,Andersen:NP_autocat:12}.  The reflection of these difficulties in CRN moment hierarchies is that moment equations at any order couple to moments at higher order,
leading to an infinite hierarchy of equations.
The standard way to deal with these is via moment-closure schemes
\cite{Schnoerr:Survey:15},
which however are ad-hoc and sometimes give unphysical results
\cite{Schnoerr:Review:17}.
In this paper, we take a different point of view. We show,
for a generic mass-action CRN with arbitrary value of $\delta$,
that the equations for the {\em factorial moments} (FM), provide
a better starting point for solving the infinite moment-hierarchy.
The structure of these equations facilitates recursively
writing FM ratios at lower-order in terms of FM ratios at higher order.
The recursions can then be solved, exactly in some cases, to
obtain any moment of interest. Our results are also  applicable to any
non-equilibrium process  describable
as a mass-action CRN, such as for example, the zero-range process
\cite{Evans:zero_range_05} \footnote{The zero range process, with periodic boundary conditions may be written as the following CRN $S_1 \rightarrow S_2 \rightarrow \cdots \rightarrow S_N \rightarrow S_1$. Here the species and complexes are the same and correspond to all the particles sitting on a site. Particle
flux into or out of the system may be easily accommodated by adding reactions of the type $S_i \xrightleftharpoons[]{}\varnothing$}.




\section{Framework and Results}

In what follows, we develop a convenient formalism for describing CRN's,
by combining a network decomposition
made standard by CRN-theory \cite{Feinberg:notes:79,Gunawardena:CRN_for_bio:03}
with the well-known stochastic process formalism due to Doi \cite{Doi:SecQuant:76}. To our knowledge, no one has combined these two methods earlier.
The description of CRN's simplifies considerably within this framework. In
addition this formalism is crucial to understanding why the equations
for the FM have the structure they do.
We hence utilize two simple examples of CRN's with non-zero deficiency, to explain both the formalism and our results.
We also provide a definition for the very important concept of deficiency.

\subsection{Two examples}

Our first example is  the following minimal model
with just one species and $\delta=1$,

\begin{figure}[ht]
  \begin{center} 
  \includegraphics[scale=0.6]{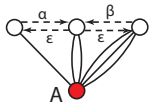}
  \caption{A compact graphical representation for a CRN: The red filled circle
    with label
    ${\rm A}$ represents the {\em species} ${\rm A}$
    while each open circle represents a {\em complex},
    such
as $2 {\rm A}$. The number of solid lines connecting each species to 
each complex, represents the {\it stoichiometry} of the complex.
Each dashed line connecting two complexes represents a 
\textit{reaction}.
    \label{fig:g_123_conn} 
  }
  \end{center}
\end{figure}

Its reaction scheme is
\begin{align}
  {\rm A} 
& \xrightleftharpoons[\epsilon]{\alpha}
  2 {\rm A} \xrightleftharpoons[\beta]{\epsilon}
  3 {\rm A}  
\label{eq:A_AA_AAA_R_scheme}
\end{align}

Another example is the following CRN with two species
and $\delta=2$:

\begin{figure}[ht]
  \begin{center} 
  \includegraphics[scale=0.6]{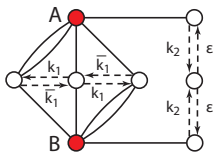}
  \caption{
 A CRN involving two species ${\rm A}$ and ${\rm B}$.
 The complex with no solid lines
 attached to it is the null node $\varnothing$. 
    \label{fig:two_species_bistable} 
  }
  \end{center}
\end{figure}

Its reaction scheme is
\begin{align}
&  {\rm B} \xrightleftharpoons[\epsilon]{k_2}
  \varnothing
 \xrightleftharpoons[k_2]{\epsilon}
      {\rm A} \nonumber \\
&      2 {\rm B} + {\rm A}
   \xrightleftharpoons[k_1]{{\bar{k}}_1}
  {\rm A} + {\rm B} \xrightleftharpoons[{\bar{k}}_1]{k_1}
  2 {\rm A} + {\rm B} 
\label{eq:cubic_2spec_scheme}
\end{align}

In the description of CRN's two matrices conventionally appear \cite{Feinberg:notes:79, Gunawardena:CRN_for_bio:03}.
An  \textit{Adjacency matrix}, denoted by ${\cal{A}}$,
is the matrix of transition rates among complexes.
The matrix element ${\cal{A}}_{ij}$ for $i \neq j$ denotes the transition
(if any) that takes complex $j$ to complex $i$, with
${\cal{A}}_{jj} \equiv -\sum_i{\cal{A}}_{ij}$.
$\cal{A}$ has, by definition,
a zero left eigenvector $\left[ 1,1,..1 \right]$.

For the network in Fig. \ref{fig:g_123_conn} the 
adjacency matrix is
\begin{equation}
  \cal{A} = 
  \left[ 
    \begin{array}{rrr}
      - \alpha & \epsilon & 0 \\
      \alpha & - 2\epsilon & \beta \\ 
       0 & \epsilon & -\beta \\ 
    \end{array}
  \right] . 
\label{eq:A_AA_AAA_rate_matrix}
\end{equation}

For the network in Fig. \ref{fig:two_species_bistable},
$\cal{A}$ is a $6 \times 6$ matrix over the
$6$ complexes: $\varnothing$, $\rm A$, $\rm B$, $\rm{A+B}$, $\rm{2A+B}$ and $\rm {A+2B}$.

\begin{equation}
  \cal{A} = 
  \left[ 
    \begin{array}{rrrrrr}
      - 2\epsilon & k_2 & k_2 & 0&0&0 \\
        \epsilon & - k_2 & 0 &0&0&0 \\ 
        \epsilon &0& - k_2 & 0 &0&0 \\ 
        0&0&0& -2 k_1& {\bar{k}}_1&{\bar{k}}_1 \\
        0&0&0& k_1& -{\bar{k}}_1& 0 \\
        0&0&0& k_1& 0 & -{\bar{k}}_1 \\
    \end{array}
  \right] . 
\label{eq:A_AA_AAA_rate_matrix}
\end{equation}

We assume mass-action rates (as in earlier work \cite{Feinberg:def_01:87, Anderson:product_dist:10}):
if $\rn_a$ is the number of particles
of species ${\rm A}$, the rate at which complex ${\rm A}$ is converted 
to any other complex is the rate constant times
$\rn_a$. Similarly the rate at which complex $2 {\rm A}$ 
takes part in any reaction is $\rn_a \left(\rn_a - 1 \right)$, {\em etc}. 

The other matrix which is useful to define is the stoichiometric matrix $Y$.
An element $y_{p,i}$ of this matrix is the amount of
species $p$ in complex $i$. We denote by $Y_p$,
the $p^{th}$ row of this matrix.

For example, in the reaction scheme of Eq. (\ref{eq:A_AA_AAA_R_scheme}), 
$Y$ is a  row vector given by

\begin{equation}
  Y  = 
  \left[ 
    \begin{array}{ccc}
      1 & 2 & 3 
    \end{array}
  \right]
\label{eq:123_y_form}
\end{equation}

For the reaction scheme of Eq. (\ref{eq:cubic_2spec_scheme}),
the $Y$ matrix is

\begin{equation}
  Y  = 
  \left[ 
    \begin{array}{cccccc}
      0 & 1 & 0 & 1 & 2 & 1 \\
      0 & 0 & 1 & 1 & 1 & 2 \\
    \end{array}
  \right]
\label{eq:cubic_y_form}
\end{equation}

where the first row $Y_1$ refers to species $\rm A$, the second row $Y_2$
to species $\rm B$ and the columns $i$ refer to the complexes
in the order mentioned above.

The time-evolution of the species in a CRN, is described
by a master equation for the probability
${\rho}_{\rn}$, where ${\rm n} \equiv \left[ {\rm n}_p \right]$
is a column vector,
with components which are  the instantaneous
numbers of the different species $p$.

For example, for the CRN of Eq. (\ref{eq:A_AA_AAA_R_scheme}), the master equation is
\begin{align}
  {\dot{\rho}}_{\rm n} 
& = 
  \left\{
    \left( e^{-\partial / \partial {\rm n}} -1 \right) 
    \left[ 
      \alpha {\rm n} + 
      \epsilon {\rm n} \left( {\rm n} - 1 \right)
    \right] 
  \right.
\nonumber \\ 
& \phantom{=} 
  \mbox{} + 
  \left. 
    \left( e^{\partial / \partial {\rm n}} -1 \right) 
    \left[ 
      \epsilon {\rm n} \left( {\rm n} - 1 \right) + 
      \beta 
      {\rm n} \left( {\rm n}-1 \right) \left( {\rm n}-2 \right)
    \right]
  \right\}
  {\rho}_{\rm n} . 
\label{eq:ME_g_123_R}
\end{align}
where the operators $e^{-\partial / \partial {\rm n}}$ (or
$e^{\partial / \partial {\rm n}}$)
act on any function $f \! \left( {\rm n} \right)$ and convert it to $f
\! \left( {\rm n}-1 \right)$ ($f \! \left( {\rm n} +1 \right)$ respectively)
\cite{Smith:LDP_SEA:11}.

For the network of Fig.~\ref{fig:two_species_bistable}, $\rm n$ becomes a
two-component index to $\rho$, which evolves under
\begin{align}
  {\dot{\rho}}_{\rm n} 
& = 
  \left\{
    \left( e^{-\partial / \partial {\rm n}_a} -1 \right) 
    \left[ 
      \epsilon + 
      k_1 {\rm n}_b {\rm n}_a
    \right] 
  \right. 
\nonumber \\ 
& \phantom{=} 
  \mbox{} + 
  \left. 
    \left( e^{\partial / \partial {\rm n}_a} -1 \right) 
    \left[ 
      k_2 {\rm n}_a + 
      {\bar{k}}_1
      {\rm n}_b {\rm n}_a \left( {\rm n}_a-1 \right) 
    \right] 
  \right.
\nonumber \\
& \phantom{=}
  \mbox{} + 
  \left. 
    \left( e^{-\partial / \partial {\rm n}_b} -1 \right) 
    \left[ 
      \epsilon + 
      k_1 {\rm n}_a {\rm n}_b
    \right] 
  \right. 
\nonumber \\
& \phantom{=} 
  \mbox{} +
  \left.  
    \left( e^{\partial / \partial {\rm n}_b} -1 \right) 
    \left[ 
      k_2 {\rm n}_b + 
      {\bar{k}}_1
      {\rm n}_a {\rm n}_b \left( {\rm n}_b-1 \right) 
    \right] 
  \right\}
  {\rho}_{\rm n}
\label{eq:ME_two_species_bistable}
\end{align}

In general, for a CRN with $P$ species,
the master equation is more conveniently written in terms
of an equation for the generating function $\phi (z) \equiv \sum_{\rn} \left(\prod_{p = 1}^P z_p^{{\rn}_p} \right) {\rho}_{\rn}$
where $z\equiv \left[ z_p \right]$ is a vector.
The generating function evolves under a \textit{Liouville equation} of
the form 
\begin{align}
  \frac{\partial \phi}{\partial \tau} = 
  - \mathcal{L} \phi
  \nonumber \\ 
  \mbox{shorthand for } \quad
  \frac{\partial}{\partial \tau} 
  \phi \! \left( z \right)
& = 
  - \mathcal{L} \! 
  \left( z , \frac{\partial}{\partial z} \right)
  \phi \! \left( z \right) . 
\label{eq:Liouville_eq_multi_arg}
\end{align}
$\mathcal{L}$ is called the \textit{Liouville Operator}.

\subsection {The Liouvillian}
$\mathcal{L}$ has a well-known representation, due to Doi \cite{Doi:SecQuant:76}, in  terms of raising and lowering operators $a^{\dagger}$ and $a$.
We provide a brief introduction to the Doi algebra below \footnote{Much more comprehensive treatments are to be found in \cite{Cardy:FTNeqSM:99,Mattis:RDQFT:98}. Interpretations of terms in the Doi algebra in the language of conventional generating functions is elaborated on in detail in \cite{Smith:evo_games:15}.}.
The Doi algebra uses the following correspondence:
\begin{align}
  z_p 
& \rightarrow a^{\dagger}_p
& 
  \frac{\partial}{\partial z_p}
& \rightarrow 
  a_p .
\label{eq:a_adag_defs}
\end{align}
It follows that the operators obey the conventional commutation algebra 
\begin{align}
  \left[ 
    a_p , a^{\dagger}_q 
  \right] = 
  {\delta}_{pq} , 
\label{eq:comm_relns}
\end{align}
where ${\delta}_{pq}$ is the Kronecker $\delta$.

Defining formal \textit{right-hand and left-hand null states},
\begin{align}
  1 
& \rightarrow 
  \left| 0 \right)
& 
  \int d^P \! z \, 
  {\delta}^P \! \left( z \right) 
& \rightarrow 
  \left( 0 \right|
\label{eq:null_states}
\end{align}
(where ${\delta}^P \! \left( z \right)$ is the Dirac $\delta$ in $P$
dimensions, and the inner product of the null states is normalized:
$\left( 0 \mid 0 \right) = 1$),
for any vector ${\rm n} \equiv \left[ {\rm n}_p \right]$, 
\begin{align}
&\prod_{p = 1}^P
  {
    a_p^{\dagger} 
  }^{{\rn}_p} 
  \left| 0 \right) \equiv 
  \left| \rn \right) .
\label{eq:number_states}
\end{align}

With these steps, the generating function $\phi$ becomes\footnote{
Note that though the generating function $\phi$ is explicitly
an analytic function of $z$,  while the state $\left|\phi \right)$ is not,
the information they carry as a power series is exactly the same. Hence, for the purpose of generating moments, the fact that both are formal power series, of $z$ in one case and of $a^{\dagger}$ in the other, suffices without worrying about convergence properties \cite{Wilf:gen_fun:06}.} 
\begin{align}
  \phi \! \left( z \right) = 
  \sum_{\rn}
  \prod_{p = 1}^P
  z_p^{{\rn}_p} 
  {\rho}_{\rn}
& \rightarrow 
  \sum_{\rn}
  {\rho}_{\rn}
  \left| \rn \right) \equiv 
  \left| \phi \right) . 
\label{eq:genfun_to_state}
\end{align}

The Liouville equation in this language
takes the form
\begin{align}
  \frac{\partial  \left| \phi \right) }{\partial \tau} = 
  - \mathcal{L} \! 
  \left( a_p , a_p^{\dagger}\right)
  \left| \phi \right) . 
\label{eq:Liouville_eq_multi_arg}
\end{align}

For example,
the Liouville operator $\mathcal{L}$ for the network
of Fig. \ref{fig:g_123_conn} (and Eq. \ref{eq:A_AA_AAA_R_scheme}) is

\begin{align}
  \mathcal{L} 
& = 
  \left( 1 - a^{\dagger} \right) 
  \left[
    \alpha a^{\dagger} a - 
    \epsilon \left( 1 - a^{\dagger} \right) 
    a^{\dagger} a^2 - 
    \beta {a^{\dagger}}^2 a^3 
  \right]
\nonumber \\ 
& = 
  \left( 1 - a^{\dagger} \right) 
  \left( a^{\dagger} \! a \right) 
  \left[
    \left( 
      \alpha - 
      \epsilon a 
    \right) + 
    \left( a^{\dagger} \! a - 1 \right) 
    \left( 
      \epsilon - 
      \beta a 
    \right) 
  \right] . 
\label{eq:L_123_R}
\end{align}

The Liouville operator corresponding to the two-species network (Fig. \ref{fig:two_species_bistable}) is 
\begin{align}
  \mathcal{L} 
& = 
  \left( 1 - a^{\dagger} \right) 
  \left[
    \left( 
      \epsilon - 
      k_2 a 
    \right) + 
    \left( b^{\dagger} \! b \right) 
    \left( a^{\dagger} \! a \right) 
    \left( 
      k_1  - 
      {\bar{k}}_1 a 
    \right) 
  \right] 
\nonumber \\
& \phantom{=} 
  \mbox{} + 
  \left( 1 - b^{\dagger} \right) 
  \left[
    \left( 
      \epsilon - 
      k_2 b 
    \right) + 
    \left( a^{\dagger} \! a \right) 
    \left( b^{\dagger} \! b \right) 
    \left( 
      k_1  - 
      {\bar{k}}_1 b 
    \right) 
  \right] . 
\label{eq:L_two_species_bistable}
\end{align}

where $\left( a^{\dagger} , a \right)$ and $\left( b^{\dagger} , b \right)$,
are creation and annihilation operators for the number components $\rn_a$ and $\rn_b$, respectively.

 

The Liouvillian may be written more compactly
in terms of the matrices $\cal{A}$ and $Y$.
To accomplish this, we need to introduce a little more notation.
Define a column vector,
\begin{align}
  {\psi}_Y^i \! \left( a \right) 
& \equiv 
  \prod_p 
  \left(a_p\right)^{y_{p,i}}
\label{eq:Psi_psi_i_def}
\end{align}

${\psi}_Y^{\dagger} \equiv {\left[ {{\psi}^{\dagger}}_Y^i \right]}^T$
is then a row vector of components defined on the indices $i$ \footnote{The index $i$ on the LHS indicates a component of the row vector and not a power.},
\begin{equation}
  {{\psi}^{\dagger}}_Y^i \! 
  \left( a^{\dagger} \right) \equiv 
  \prod_p 
  \left({a^{\dagger}}_p\right)^{y_{p,i}}
\label{eq:psi_dag_i_def}
\end{equation}

For example, for the two-species network, these
are simply

\begin{align}
  {\psi}^{\dagger} 
& = 
  \left[
    \begin{array}{cccccc}
      1 & a^{\dagger} & b^{\dagger} & 
      a^{\dagger} b^{\dagger} & 
      {a^{\dagger}}^2 b^{\dagger} & 
      a^{\dagger} {b^{\dagger}}^2 
    \end{array}
  \right]
\nonumber \\
  {\left( \psi \right)}^T
& = 
  \left[
    \begin{array}{cccccc}
      1 & a & b & ab & a^2 b & ab^2 
    \end{array}
  \right]
\label{eq:num_vecs_two_species_bistable}
\end{align}

In this formalism, the Liouville operator takes on the simple form,
\begin{align}
  - \mathcal{L} 
& = 
  {\psi}^{\dagger}_Y 
  {\cal A}
  {\psi}_Y
\label{eq:L_psi_from_A}
\end{align}

\subsection {The Moment Hierarchy}
Entirely equivalent to solving the master equation or the Liouville equation, is to
solve the \textit{moment hierarchy}, namely to compute the time-dependent 
values of \textit{all} the relevant moments in the problem.
The set of equations for all these moments, obtained directly
from the master equation or the Liouville equation, 
is referred to as the moment hierarchy, because
usually lower-order moments couple to higher-order ones, resulting in an infinite
hierarchy of equations. Solving the moment hierarchy  is hence by no means
a simple task and often involves making approximations. In what follows, we
demonstrate that for any mass-action CRN, the equations for
the \textit{factorial moments} (rather than the equations for ordinary moments) take on 
a particularly tractable form. For the examples we consider, we show how this
tractability helps in solving the entire moment hierarchy in the steady state.

In order to see this, we first need to write down the 
equations for the moments.
The time dependence of arbitrary moments is easily extracted from the
Liouville equation via the Glauber inner product,
which is a standard construction \cite{Cardy:FTNeqSM:99,Mattis:RDQFT:98}. 
In the interest of completeness, we provide all relevant details in what follows.
As mentioned earlier, we will prefer instead to look
at the {\it factorial} moments (FM). In order to define these,
consider, for a single component ${\rn}_p$ and power $k_p$,  the quantity
\begin{align}
  {\rn}_p^{\underline{k_p}} 
& \equiv 
  \frac{
    {\rn}_p !
  }{
    \left( {\rn}_p - k_p \right) ! 
  }
& ; \; k_p \le {\rn}_p 
\nonumber \\ 
& \equiv 
 0 
& ; \; k_p > {\rn}_p .
\label{eq:factorial_moment_not}
\end{align}

For a vector $k \equiv \left[ k_p \right]$ of powers and a vector $\rn$ of
instanstaneous numbers of the species, we
introduce the \textit{factorial moment} indexed by $k$, as
the expectation
\begin{equation}
  \left< 
    \rn^{\underline k}
  \right>   \equiv
  \left<
    \prod_p
    {\rn}_p^{\underline{k_p}}
  \right> . 
\label{eq:Phi_def}
\end{equation}

The FM are generated by the action of the lowering operator on the number state.
In particular, for any non-negative integer $k_p$,
\begin{align}
  a_p^{k_p}  
  \left| \rn \right)
& = 
  {\rn}_p^{\underline{k_p}}
  \left| \rn - k_p \right) 
\nonumber \\ 
  {a^{\dagger}_p}^{k_p} a_p^{k_p} 
  \left| \rn \right)
& = 
  {\rn}_p^{\underline{k}}
  \left| \rn \right) .
\label{eq:trunc_fact_extract}
\end{align}
where $\left| \rn - k_p \right)$ is the number state with $k_p$
subtracted from ${\rn}_p$ and all ${\rn}_q$ for $q \neq p$ unchanged. 

The time dependence of the FM is then simply given by
\begin{align}
  \frac{\partial}{\partial \tau}
  \left< 
    \prod_p 
    {\rn}_p^{\underline{k_p}}
  \right> \equiv  \frac{\partial}{\partial \tau}
  \left< 
    \rn^{\underline k}
  \right> 
& = 
  \left( 0 \right| 
  e^{\sum_q a_q}
  {
   \prod_p \left(a_p \right)
  }^{k_p}
  \left( - \mathcal{L} \right)
  \left| \phi \right) 
\nonumber \\ 
& = 
  \left( 0 \right| 
  e^{\sum_q a_q}
  {  \prod_p \left(a_p \right)
  }^{k_p}
  {\psi}^{\dagger}_Y \! \left( a^{\dagger} \right) 
  {\cal A} \, 
  {\psi}_Y \! \left( a \right) 
  \left| \phi \right) 
\label{eq:fac_mom}
\end{align}

In writing Eq. (\ref{eq:fac_mom}), the fact that
all number states are normalized with respect to the \textit{Glauber
inner product}, defined by 
\begin{align}
  \left( 0 \right|
  e^{\sum_p a_p}
  \left| \rn \right) = 1 , \quad
  \forall \rn 
\label{eq:Glauber_inn_prod}
\end{align}
is used.

The Glauber inner product with a generating function is simply the
trace of the underlying probability density: 
\begin{align}
  \left( 0 \right|
  e^{\sum_p a_p}
  \left| \phi \right) = 
  \sum_{\rn}
  {\rho}_{\rn} = 1 . 
\label{eq:Glauber_is_trace}
\end{align}


Eq. (\ref{eq:fac_mom}) denotes the time-evolution of a generic
FM for a CRN with an arbitrary number of species.
In particular, the equation for the first moment
takes on a simple form. Note that a
first moment, for a CRN of $P$ species, is 
specified by a vector $k \equiv \left[ k_p \right]$,
with only one of the $k_p$'s being non-zero (and having the value $1$). This corresponds
 to computing the average value of the number of one specific species $p$.
In this case, from Eq. (\ref{eq:fac_mom}), we need
only to commute $a_p$ through $ {\psi}^{\dagger}_Y \! \left( a^{\dagger} \right)$ \footnote{The general closed form expression for the commutation of $a_p^{k_p}$ through any power of $a^{\dagger}$ is given in Eq. (\ref{eq:comm_rules}).} to obtain,

\begin{align}
\frac{\partial}{\partial \tau} \left< {\rm n_p} \right> &= 
  Y_p {\cal A} \, 
    \left( 0 \right| 
    e^{\sum_q a_q} {\psi}_Y \! \left( a \right)   \left| \phi \right) 
\label{eq:CRN_std_rate_eqn}
\end{align}
In what follows, we refer to the inner product in
Eq.~(\ref{eq:CRN_std_rate_eqn})
as $E\left[{\psi}_Y \! \left( a \right) \right]$ to simplify notation.
Note that, using the above definitions, $E\left[a_p^{k_p}\right] \equiv \langle n_p^{\underline {k_p}} \rangle$: the FM of order $k_p$.


\subsection{Deficiency}

It is useful at this stage to
understand the relations between the dimensions of the
matrix ${\cal A}$ and the matrix $Y{\cal A}$ (or the row vector $Y_p{\cal A}$).
The reason for considering these is that, as we see from Eq.~(\ref{eq:CRN_std_rate_eqn}), all steady states must lie in $ \ker\left( Y{\cal A} \right)$ (since
the steady state condition implies that the
LHS of Eq.~(\ref{eq:CRN_std_rate_eqn}) must vanish).
This can happen either because the steady state lies in $\ker {\cal A}$
(and so  vanishes directly by the action of ${\cal A}$) or because
the steady state \textit{does not} lie in $\ker {\cal A}$
but nevertheless lies in $\ker\left(Y {\cal A} \right)$. The difference between
these two situations, as we will see, summarises the difference between
$\delta=0$ and $\delta \neq 0$- networks.

By definition, since the number of columns (and rows) in the
matrix ${\cal A}$ is equal to the number of complexes ${\mathcal{C}}$, matrix
${\cal A}$ has dimension $ {\mathcal{C}} $. Then from elementary considerations,
\begin{align}
  \dim \left({\cal A} \right) &= 
  \dim \left( \Ima \left( {\cal A} \right) \right) + 
  \dim \left( \ker \left( {\cal A} \right) \right) \nonumber \\
& = 
  \dim \left( \Ima \left( Y {\cal A} \right) \right) + 
  \dim 
  \left( 
    \ker Y \cap \Ima \left( {\cal A} \right) 
  \right) \nonumber + l \\
& \equiv
  s + \delta + l. 
  \label{eq:A_ident}
\end{align}

Here $ \dim \left( \ker \left( {\cal A} \right) \right) \equiv l$
where $l$ is the number of linkage classes \footnote{A linkage class is a connected component of the directed graph representing the CRN; $l=1$ for the CRN described by Eq. (\ref{eq:A_AA_AAA_R_scheme}) and $l=2$ for the CRN described by Eq. (\ref{eq:cubic_2spec_scheme})}. In Eq. (\ref{eq:A_ident}), the $  \dim \left( \Ima \left( {\cal A} \right) \right) $ is further split into those vectors
that either lie \textit{both} in the $\Ima ({\cal A})$ and $ \ker \left( Y \right)$ or lie in the  $\Ima \left( Y {\cal A} \right)$. 

Eq. (\ref{eq:A_ident})
provides a definition for the parameter $s$ and deficiency $\delta={\mathcal{C}} - s - l$ \cite{Feinberg:def_01:87}.
For the CRN in Fig. ~\ref{fig:g_123_conn}, $\mathcal{C}=3,l=1,s=1$ giving $\delta =1$.
For the CRN in Fig.  ~\ref{fig:two_species_bistable}, $\mathcal{C}=6,l=2,s=2$ giving $\delta =2$.



For $\delta =0$ networks, all steady states
lie simultaneously in $\ker \left({\cal A} \right)$ and in $\ker \left( Y {\cal A} \right)$ and are termed \textit{complex-balanced}.
If $\delta > 0$, this is no longer true.
In what follows, we derive some new
results for CRN's in this category.

From the above discussion, it follows that
we can define basis vectors
${\left\{ e_i \right\}}_{i = 1}^s$ for ${\ker \left(Y{\mathcal{A}} \right)}^{\perp}$, the space of vectors perpendicular to those lying in $\ker \left( Y {\cal A} \right)$.

Let also ${\left\{ {\tilde{e}}_j \right\}}_{j = 1}^{\delta}$ be a
basis for $\ker \left( Y{\mathcal{A}} \right) / \ker \left({\mathcal{A}} \right)$, the space of vectors lying in $\ker \left( Y {\cal A} \right)$ but {\em not} in $\ker \left( {\cal A} \right)$. 

 It follows that jointly $\left\{ {\left\{ e_i \right\}}_{i =
1}^s, {\left\{ {\tilde{e}}_j \right\}}_{j = 1}^{\delta} \right\}$ form
a basis for ${\ker \left( {\mathcal{A}} \right)}^{\perp}$.

Then from Eq.~(\ref{eq:L_psi_from_A}),
\begin{align}
  - \mathcal{L} 
& = 
  {\psi}^{\dagger}_Y 
  {\mathcal{A}}
  \left\{ 
    \sum_{i = 1}^s
    e_i e_i^T + 
    \sum_{j = 1}^{\delta}
    {\tilde{e}}_j {\tilde{e}}_j^T 
  \right\} 
  {\psi}_Y
\label{eq:Ls_reps}
\end{align}




That the second sum does not play a role for $\delta =0$ networks,
has implications for the steady state, as we will see in Section II F.
It is easy to explicitly work out these basis vectors for specific
examples (such as the networks of Fig. \ref{fig:g_123_conn}
and Fig. \ref{fig:two_species_bistable}) \cite{Smith:LP_CRN:17}.

\subsection{Equation for the Factorial Moments}
Eq. (\ref{eq:fac_mom}) is valid for a generic FM, but may be simplified
further by writing the RHS in terms of the matrices $Y$ and ${\cal A}$,
in correspondence to the equation for the first moment Eq. (\ref{eq:CRN_std_rate_eqn}).
In order to see this, we need to understand what terms
we get when we commute $a_p^{k_p}$ through $ {\psi}^{\dagger}_Y \! \left( a^{\dagger} \right)$ 
which contains terms like ${a_p^{\dagger y}}$.
For non-negative integers $k_p$ and $y$, we can use the relation 
\begin{align}
  {\left( a_p^{k_p} {a_p^{\dagger y}} \right)}
&  =   \sum_{j = 0}^{k_p}
  \left(
    \begin{array}{c}
      {k_p} \\ j 
    \end{array}
  \right)
  {
    {y}^{\underline j} {a_p^{\dagger}}^{y-j} a_p^{k_p-j}
  } 
\label{eq:comm_rules}  
\end{align}  
where $y^{\underline 0}=1$. For $j \neq 0$, 
${y}^ {\underline j} = (y)(y-1) \cdots (y-j+1)$ for $j-1 \leq y$
and ${y}^{\underline j} =0$ otherwise. It is now easily seen that,
\begin{align}
  \left( 0 \right|
  e^a 
  {
    \left( a_p \right)
  }^{k_p} 
  {{\psi}^{\dagger}}_Y^i \! \left( a^{\dagger} \right) 
= 
  \left( 0 \right|
  \sum_{j = 0}^{k_p}
  \left(
    \begin{array}{c}
      k_p \\ j 
    \end{array}
  \right)
  {
    \left( Y_{p} \right) 
  }^{\underline j} 
  e^a 
  {
    \left( 
      a_p 
    \right) 
  }^{k_p - j}
\label{eq:np_psii_comm_exp}
\end{align}
where $\left( Y_{p} \right)^{\underline j} $ is the matrix $Y$ with
the elements in the $p^{th}$ row modified to $y_{p,i}^{\underline j}$.  
$\psi \left( a^{\dagger} =1 \right) = {1}$, because $
\left( 0 \right| e^a a^{\dagger y} = \left( 0 \right| {\left( 1+
a^{\dagger} \right) }^y e^a = \left( 0 \right|e^a $.

The equation for the time-dependence of  $\rn_p^{\underline {k_p}}$
may hence be compactly written as
\begin{align}
  \frac{\partial}{\partial \tau}
  \left< 
    \rn_p^{\underline {k_p}}
  \right> 
& = 
  \sum_{j = 0}^{k_p}
  \left(
    \begin{array}{c}
      k_p \\ j 
    \end{array}
  \right)
  {
    \left( Y_p \right) 
  }^{\underline j} {\cal A}
  E\left[ a_p^{k_p-j} {\psi}_{Y} \! \left( a \right)\right]
\label{eq:fac_mom1}
\end{align}



Note that in Eq. (\ref{eq:fac_mom1}),
$\underline{j}=0$ does not contribute since this multiplies $\cal{A}$
by a row vector of $1$'s, which is a zero left eigenvector.
Hence for $k_p=1$, only the $j=1$ term contributes. This
gives $\left( Y_{p} \right)^{\underline 1} = Y_p $, resulting
in the RHS of Eq. (\ref{eq:CRN_std_rate_eqn}).


Eq.(\ref{eq:fac_mom1}) may also be easily generalized in order to calculate mixed moments as in Eq. (\ref{eq:Phi_def}). For this
we need to  consider the action of the lowering operators
$\prod_p a_p^{k_p}$ (as in Eq. \ref{eq:fac_mom})  which, by their action on
$\left| \phi \right) $
result in mixed moments $\left< \prod_p {\rn}_p^{\underline{k_p}}  \right>$.
 By considering the generalisation of Eq. (\ref{eq:comm_rules}),
the equation for the time derivative of such a mixed moment is seen to be,
\begin{align}
  \frac{\partial}{\partial \tau}
  \left< 
    \prod_p 
    {\rn}_p^{\underline{k_p}}
  \right> 
& = 
  \sum_{j_1 = 0}^{k_1}
  \left(
    \begin{array}{c}
      k_1 \\ j_1 
    \end{array}
  \right) \ldots 
  \sum_{j_P = 0}^{k_P}
  \left(
    \begin{array}{c}
      k_P \\ j_P 
    \end{array}
  \right)  
  \left[ 
    \dot{\prod_p}
    Y_p^{\underline{j_p}} \,  
  \right]
  {\cal A}
  E \left[
    {\Psi}_{Y + \left( k - j \right)} \! \left( \rn \right) 
  \right]
\nonumber \\ 
\label{eq:Glauber_moment_fact_prod}
\end{align}

The notation ${\dot{\prod}}_p$ denotes a product over species $p$
within each index $i$ of the row vectors $Y_p^{\underline{j_p}}$.
Note that in the sums over $j_p$, we must now retain the $j_p = 0$
entries, because even if one index $Y_p^{\underline{j_p}} = {\left[ 1
\right]}^T$, there may be others in the sum where $j_{p^{\prime}} \neq
0$, and the product $\left( {\dot{\prod}}_p Y_p^{\underline{j_p}}
\right)   {\cal A} $ is only assured to vanish when all $j_p = 0$.
The term $ E \left[ {\psi}_{Y + \left( k - j \right)} \! \left( \rn \right)\right]$
is a shorthand notation for 
$ E\left[\prod_p a_p^{k_p-j_p} {\psi}_{Y} \! \left( a \right)\right]$.

Note that though Eq. (\ref{eq:fac_mom1}) and Eq. (\ref{eq:Glauber_moment_fact_prod}) may be  derived directly
from the master equation (without going through the Doi algebra),
the simplification that comes from noting the relation of
the coefficients to the matrices $Y$ and ${\cal A}$
is only possible within the formalism we introduce here.
This in turn helps in writing and solving
recursion relations to solve the entire moment hierarchy as we demonstrate
in Section II G.

\subsection{A one-line proof of the ACK theorem}
We explain how the ACK theorem follows very simply from
the considerations above. Without loss of generality we
limit this discussion to Eq. (\ref{eq:fac_mom1}), for
ease of presentation.

From the considerations in Section II D,
Eq. (\ref{eq:fac_mom1}) may be re-written as,
\begin{align}
  \frac{\partial}{\partial \tau}
  \left< 
    \rn_p^{\underline k_p}
  \right> 
  & = 
  \sum_{j = 0}^{k_p}
  \left(
    \begin{array}{c}
      k_p \\ j 
    \end{array}
  \right)
  {
    \left( Y_p \right) 
  }^{\underline j} {\mathcal{A}}
    \left\{ 
    \sum_{i = 1}^s
    e_i e_i^T + 
    \sum_{j = 1}^{\delta}
    {\tilde{e}}_j {\tilde{e}}_j^T 
    \right\}
    \nonumber \\
&  \mbox{} 
  \times E \left[ a_p^{k_p-j} {\psi}_{Y} \! \left( a \right) \right]
\label{eq:fac_mom3}
\end{align}

In particular, the equation for the first moment
Eq. (\ref{eq:CRN_std_rate_eqn}), can be written as
\begin{align}
  \frac{\partial}{\partial \tau}
  \left< 
    {\rn_p}
  \right> 
  & = 
  {
    Y_p {\cal A}
  \sum_{i = 1}^s
  e_i e_i^T 
  E \left[ {\psi}_{Y} \! \left(a\right) \right]}
\label{eq:fac_mom4}
\end{align}
where we have used the fact that all other basis
vectors are projected to zero by $Y_p {\cal A}$. 

Hence for $\delta=0$ networks with mass-action
rates,
the entire hierarchy of moments, Eq. (\ref{eq:fac_mom3}) for any value of $k_p$, 
is satisfied
if
\begin{equation}
 e_i^T E \left[\left[{\psi}_{Y} \! \left( a \right)\right]\right] =0
\label{eq:projectors_cond}
\end{equation}
for every $i=1,\cdots s$. The notation $E\left[\left[ \,\,
    \right]\right]$ denotes
now an average over a specific distribution:  a Poisson distribution. 
Note that for a Poisson distribution,
an equation for the first moment is the same as a {\it rate equation} 
(since $\langle \rn^{\underline k} \rangle = \langle \rn \rangle ^k$).
Hence, for $\delta =0$ networks,
the condition that the rate equation has a unique solution
also guarantees that the entire moment hierarchy is solved,
whereby follows the ACK theorem \cite{Anderson:product_dist:10}.


\subsection{Steady-state Recursions}
For CRN's for which  $\delta \neq 0$, there is no general way
to satisfy the full moment-hierarchy
of Eq. (\ref{eq:Glauber_moment_fact_prod}) by demanding that any combination of
$e$ and $\tilde{e}$ vanish.

Note though that the sums $\sum_j$ in  Eq. (\ref{eq:Glauber_moment_fact_prod}), 
only extend from $j=0$ to $j=j_{\rm max}$ with
the latter determined by when row
$Y_p^{\underline j_{\rm max}}$ vanishes. For the CRN in
Fig. \ref{fig:g_123_conn}, $j_{\rm max} =4$,
while for the CRN in Fig. \ref{fig:two_species_bistable}, there are two sums over $j$
in Eq. (\ref{eq:Glauber_moment_fact_prod}), both with  $j_{\rm max} =3$.


This  helps us write Eq. (\ref{eq:fac_mom1}) (or Eq. (\ref{eq:Glauber_moment_fact_prod}) in the general case), as  a recursion relation
for the ratios of FM's in the steady state.
We demonstrate this for the two examples introduced above. For the CRN of Fig. \ref{fig:g_123_conn}
if we define $  R_k \equiv \frac{\left<{\rn}^{\underline k}\right>}{\left<{\rn}^{\underline {k-1}}\right>} $\footnote{for one species $k=k_p$}, 
then Eq. (\ref{eq:fac_mom1}) may be rewritten {\it exactly} as the
following recursion relation for $k>1$\footnote{The moment recursions for this CRN leave $k=1$ undertermined. However this does {\em not} mean that $R_1$ is free to take any value. Moments of a probability distribution satisfy inequalities \cite{vanKampen:Stoch_Proc:07} such as the elementary relation $R_2 \geq (R_1-1)$. These presumably constrain the first moment to its actual value.},
\begin{equation}
  R_k=  \frac{(k-1)\left(\frac{\alpha}{\beta}+ \frac{\epsilon}{\beta}(k-2)\right)} {(k-1) \left(2 R_{k+1} + (k-2) -\frac{\epsilon}{\beta}\right) + R_{k+2}R_{k+1} - \frac{\alpha}{\beta} }
\label{eq:rec_123}  
\end{equation}

\begin{figure}
  \begin{center}
  \includegraphics[scale=0.2,angle=270]{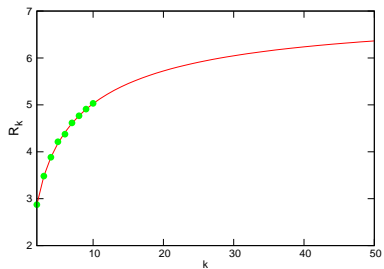}
  \caption{Numerical evaluation of the recursions of Eq. \ref{eq:rec_123} (line) compared to values from Monte-Carlo simulations (symbols) for $\alpha=100$, $\beta=10$ and $\epsilon =70$. For large $k$, $R_k$ saturates to $\epsilon/\beta = 7$ as explained in the text.
    \label{fig:rec_123_comp} 
  }
  \end{center}
\end{figure}

We have written the recursion for $R_k$ for descending $k$ because,
while we do not know the value of $R_k$ for
small $k$, we do know it for large $k$, where $R_k \sim \epsilon/\beta$
(as evident from  Eq. \ref{eq:rec_123}).
If we begin from this `asymptotic' value
at arbitrarily large $k$, we have a procedure to
obtain the value of $R_k$ all the way down to $k=2$, for any choice of
parameters \footnote{We often want to obtain the actual moments
  and not just their ratios. Note that this is possible
  since $R_1 \equiv \left<n\right>$. Hence, $\langle n^{\underline{2}} \rangle = R_2R_1$; $\langle n^{\underline{3}} \rangle = R_3R_2R_1$ {\em etc}.}
The result is shown in Fig. \ref{fig:rec_123_comp}. Eq. (\ref{eq:rec_123}) being 
\textit{exact }, the results of the recursions and the Monte-Carlo simulations
agree to arbitrary accuracy, limited only by the amount of averaging done in the simulations (and we expect this to be the case for
any set of parameters)\footnote{The downward recursion Eq. (\ref{eq:rec_123})
  may however not converge below $k$-values much smaller than $\left<n\right>$ for parameter values which make the latter large. In this case,
an \textit{upward} recursion, for larger $k$ in terms of smaller $k$ can be written and both
upward and downward recursions solved simultaneously. In \cite{Smith:LP_CRN:17}, we elaborate on this further.}.

The one-species case we have considered is an example of a birth-death process
\cite{vanKampen:Stoch_Proc:07} for which many results are known, including the steady state. The recursions Eq. (\ref{eq:rec_123}) however, give us a particularly
easy way, albeit numerical, to obtain this steady state.
In addition, while there exists no general formalism to obtain the steady state for CRN's which are {\em not} birth-death processes, the above procedure is, in principle, applicable to any CRN, such as the two-species CRN  of Fig. \ref{fig:two_species_bistable}, as we show below.

The moment hierarchy for the two-species case 
consists of mixed moments such as
$ \langle {\rm n}_a^{\underline k}{\rm n}_b^{\underline k^{\prime}} \rangle$.
This CRN has no conservation
law, so solving the full moment hierarchy is equivalent to solving for the full probability distribtion $P\! \left(\rn_a,\rn_b \right)$ which is, in addition, {\em not} factorized. In analogy with the one-species case,
we can write a coupled
set of recursions for the quantities $T_k \equiv \frac{ \langle {\rm n}_a^{\underline k}{\rm n}_b^{\underline k} \rangle}{\langle {\rm n}_a^{\underline{k-1}}{\rm n}_b^{\underline{k-1}}\rangle}$ and $S_k \equiv \frac{\langle {\rm n}_a^{\underline k}{\rm n}_b^{\underline{k-1}}\rangle }{\langle {\rm n}_a^{\underline{k-1}}{\rm n}_b^{\underline{k-1}}\rangle}$. For large $k$, the equations for the FM predict that $T_k \sim \left(k_1/{\bar k}_1\right)^2$ and $S_k \sim \left(k_1/{\bar k}_1\right) $.

\begin{figure}
  \begin{center}
   \includegraphics[scale=0.2,angle=270]{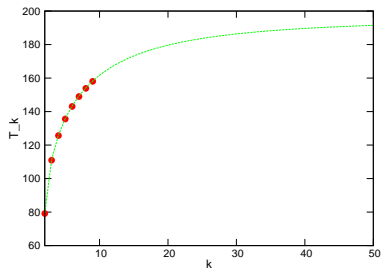}
  \caption{Numerical evaluation of the recursions of Eq. \ref{eq:rec_two_species} (line) compared to values from Monte-Carlo simulations (symbols) for $k_1=14$, $\bar{k}_1=1$,$\epsilon =36$ and $k_2=49$. For large $k$, $T_k$ saturates to $\left(k_1/{\bar k}_1\right)^2 = 196$. {The values obtained from the recursions and simulations agree upto the first decimal place for any $k$.}
    \label{fig:rec_two_species} 
  }
  \end{center}
\end{figure}

Using the symmetries of this CRN (in the exchangeability of the
species $\rm A$ and $\rm B$; hence $\langle {\rm n}_a^{\underline k}{\rm n}_b^{\underline k^{\prime}} \rangle = \langle {\rm n}_a^{\underline k^{\prime}}{\rm n}_b^{\underline k} \rangle$), and approximating $\frac{\langle {\rm n}_a^{\underline {k-2}}{\rm n}_b^{\underline{k}}\rangle }{\langle {\rm n}_a^{\underline{k-1}}{\rm n}_b^{\underline{k-1}}\rangle} \sim 1$\footnote{We have verified this numerically. A
  theoretical justification comes from looking at  the analytic form of the FM in the large-$k,k^{\prime}$ limit \cite{Smith:LP_CRN:17}. We can show that to leading order the FM are only functions of $k+k^{\prime}$ thus validating this approximation.}, 
we obtain two coupled recursions,
\begin{align}
T_k &= \frac{2\epsilon k -\epsilon + k k_1(k-1)(2k-3)}{S_{k+1}C + T_{k+1}D + E} \nonumber \\
S_k &= T_k\left({S_{k+1}C_1 + T_{k+1}D_1 + E_1} \right)   
\label{eq:rec_two_species}  
\end{align}  
where $C$, $D$ {\em etc} are functions of $k$ as well as the rate constants
$k_1$, ${\bar{k}_1}$ {\em etc} in the problem. Again, for large $k$, it is easy to see from the recursions (after putting in the expressions for $C$, $D$ {\em etc}), that $T_k \sim \left( \frac{k_1}{\bar{k}_1} \right)^2$ and $S_k \sim \frac{k_1}{\bar{k}_1}$ as required by the equations for the FM. Beginning from this value at some
arbitrarily large value of $k$, we can predict values for $T_k$ all the
way down to $k=2$ as shown in Fig. \ref{fig:rec_two_species}.

Note that $R_k$ saturating to a constant value independent of $k$ in the one-species case is as if the large-$k$ moments obey a Poisson distribution
with parameter $ \epsilon/\beta$ \footnote{$R_k$ could also be a constant
  if the distribution was a delta function around $\epsilon/\beta$.
  This can however not happen when fluctuations in the number are possible.}.
Similarly $T_k$ and  $S_k$ saturating to constant values is equivalent to
the large-$k$ behaviour of the two-species system being describable
by a factorised Poisson distribution with parameter $\left(k_1/{\bar k}_1\right)$. From the form of Eq. (\ref{eq:fac_mom1}) in the steady state,
it is evident that, even with an arbitrary number of species,
there will always be a limited number of terms which will dominate
for large moments. Demanding that these terms vanish will hence always lead
to a factorized Poisson distribution which will approximately (up to corrections
of order $1/k$) solve the moment hierarchy.
On the other side, at $k=1$, the equation for the
first moment can also be solved by postulating a factorized Poisson distribution
with the parameter of the Poisson determined by the rate equation of the problem.
These two Poisson distributions have different parameters and are both,
for a $\delta \neq 0$- network, only approximations for the true distribution.
Nevertheless, they are helpful in implementing a systematic
approximation procedure to solve the moment hierarchy as we elaborate in a
following paper \cite{Smith:LP_CRN:17}.

\subsection{Quasi Steady States}

The CRN's we have considered so far have been {\em reversible}
in the sense that every reaction is accompanied by its reverse.
We now consider a CRN which is neither reversible nor even weakly reversible:

\begin{align}
&  {\rm B} \xrightarrow {\beta} {\rm A} \nonumber \\
&     {\rm B} + {\rm A}
   \xrightarrow {\alpha} {2 \rm B} .
\label{eq:quasist}
\end{align}

This CRN has been considered in \cite{Anderson:ACR:14} in the context of
understanding properties of the quasi-stationary distribution. The
true steady state of this model is an absorbing state with $\rn_b=0$.
However when $\rn_a+ \rn_b =\rm  M$, and $\rm M$ is large, the
system could take a very long time to reach this absorbing state,
and reach instead a quasi-stationary distribution.
All properties of the quasi-stationary distribution are easily
derivable for this model \cite {Anderson:ACR:14} and it
is seen that as $\rm M \rightarrow \infty$, this
distribution is a Poisson with parameter
$\beta/\alpha$ \cite {Anderson:ACR:14}.
The equations for the FM, give this result very easily as well.
If we define $ X_k \equiv \frac{\left<{\rn_a}^{\underline k}{\rn_b}^{\underline {k^{\prime}-1}}\right>}{\left<{\rn_a}^{\underline {k-1}} {\rn_b}^{\underline {k^{\prime}-1}}\right>}$, then it is easily seen that the  CRN of Eq. (\ref{eq:quasist}) leads to the
recursions
\begin{equation}
X_k = \frac{k\beta {\rm M}^2}{k\alpha {\rm M}^2 + A + X_{k+1}B + X_{k+1}X_{k+2} \alpha (k+k^{\prime})}
\end{equation}  
$A$ and $B$ are linear in $\rm M$, and hence for large $\rm M$ and $k^{\prime}=1$, $X_k \sim \frac{\beta}{\alpha}$ for any $k$, as expected for the ratios of the
FM of a Poisson distribution \footnote{Note that $X_k=\rm M$ is also a solution for $k^{\prime}=1$. This is the absorbing state.}.

\section{Conclusion}
To conclude, the structure of the equations for the FM
help us write them as recursions for ratios of FM, which
then can be solved numerically,
beginning from an asymptotic estimate (predicted by the equations themselves).
The equations for the FM (Eq. \ref{eq:fac_mom1} or Eq. \ref{eq:Glauber_moment_fact_prod}) are exact and given any CRN,  are easy to write down.
In this paper, we have illustrated this procedure  for two
toy models. However, there are several physically
relevant model-CRN's in the biochemistry, systems-biology, ecology
and epidemiology contexts,
to which we expect to be able to apply our methods.

It should be noted however, that
except in the case of very few species, or very simple stoichiometry,
the recursions obtained from these equations could
get complicated to solve. It would hence be very useful if this
procedure could be systematised in some way independent
of the particular CRN under study, perhaps
with the help of some of the techniques available
in the large body of work that exists on efficient
ways to truncate the moment hierarchy in CRN's \cite{Schnoerr:Survey:15}. 
In \cite{Smith:LP_CRN:17}, we have provided alternate
approximation schemes (differing from moment-closure schemes)
for the FM equations, related to asymptotic expansions in the
low-$k$ and large-$k$ limit. These methods
might be applicable, even in the case when recursions like
Eq. (\ref{eq:rec_123}) and Eq. (\ref{eq:rec_two_species}) are hard
to obtain for  CRN's with many species.

CRN's with non-mass-action kinetics
could also be interesting to look at \cite{Anderson:non_mass_action:16}.
Finally, though we have only concentrated on the
static properties here, the Liouvillian contains all 
information on the dynamics as well, which can be investigated further,
in the spirit of \cite{Polettini:diss_def:15}.




{\it Acknowledgements}: SK would like to thank Artur Wachtel
for very useful discussions during the Nordita
program `Stochastic thermodynamics in biology' (2015).  DES thanks
Nathaniel Virgo for discussions and the Stockholm University Physics Department for hospitality 
while this work was being carried out. DES acknowledges support from NASA Astrobiology Institute Cycle 7 Cooperative Agreement Notice (CAN-7) award:
Reliving the History of Life: Experimental Evolution of Major Transitions.


\bibliography{DES}

\end{document}